# Influence of Facebook in Academic Performance of Sri Lankan University Students

Thuseethan, S.[1], Kuhanesan, S.[2]

*Department of Computing and Information Systems, Sabaragamuwa University of Sri Lanka*
*thuseethan@gmail.com*
*Department of Physical Science, Vavuniya Campus of the University of Jaffna*
*kuhan9@yahoo.com*

**Abstract: Facebook is only an electronic communication between human but unfortunately it has become an addiction for all. This paper examines the usage of Facebook among university students and its influence in their academic performance. The impact of Facebook can either be good or bad on university students and in their academic activities. Even though a closer look on the real impact of Facebook reveals that it leads to several problems in university students' academic performances. Today Facebook is somehow destroying the future and academic carrier of university students. At the same time also intended to find the significance of use of Facebook by University students in their academic success with the help of a survey conducted to collect the data among more than 250 students of different Universities in Sri Lanka.**

## I. INTRODUCTION

Involvement of technology aspects for social needs has become the major communication strategy for most people in the past decade. Internet is an inseparable part of human life and it involves in variety of routine activities. Social media has revolutionized corporate communications, rapidly changing the way that public relations campaigns or programs are distributed and measured (Matthews, L., 2010). Social networking has taken the world towards a rich communication society. Facebook, Twitter, LinkedIn and MySpace are the most popular social networking sites providing the electronic communication within society. In the last five years the rapid growth of social media that has been observed is indicative of its importance and its integration into the daily lives of many people in Sri Lanka (Thuseethan, S. and Vasanthapriyan, S., 2014).

Facebook is at the leader of the social media race with more active users' worldwide. It became one of the most frequently accessed website at the moment. Facebook founded in 2004 by Mark Zuckerberg, Eduardo Saverin, Dustin Moskovitz, and Chris Hughes, who are alumni of Harvard. The typical University culture loves Facebook deeply, builds the lifestyle, rather than just a hobby or a fun time passing activity. Academic success is supreme goal to any student, with the social and family responsibility they have. University students are one the major group using Facebook for fun, with the main purposes as connect with their friends, sharing day to day activities, using features such as photo sharing, publishing wall posts, and stating their status updates.

Because of the social media platform's widespread adoption by college students, there is a great deal of interest in how Facebook use is related to academic performance (Junco, R., 2012). Like other social networking sites Facebook is severely destroying the academic life of university students. In 2008, half of the students were completely unfamiliar with Facebook, while in 2009 all our respondents were aware of it and 59% of them were also using it on a regular basis (Nicola Cavalli, Elisabetta Ida Costa, Paolo Ferri, Andrea Mangiatordi, 2011). Facebook use is nearly ubiquitous among U. S. college students with over 90% active participation among undergraduates (N. Ellison, C. Steinfeld, and C. Lampe, 2011). Even in the developed countries Facebook is widely access by students.

This paper reviews the influence of Facebook in university students' academic activities and further analyzes both positive and negative impacts of using Facebook.

## II. BACKGROUND OF THE STUDY

### A. Social Networking Sites

Social media consists of online technologies, practicing activities or societies that people use to generate content and share thoughts, visions, experiences and viewpoints with each other (Television Bureau of Advertising, Inc., 2009). The word social networking is known as the alliance of individuals into specific set of potential groups or subdivisions. Social networking allows individuals to express their thoughts to other users. Social networking is the leader in promoting digital journalism (Thuseethan, S. & Vasanthapriyan, S., 2014). Social network used for several purposes like promoting or distributing the news contents all over the world.

Social networking sites and Facebook socializing via the internet has become an increasingly important part of young adult life (Gemmill & Peterson, 2006). Most of the high schools, colleges and universities get connected by internet encompass individuals who are looking forward to mingle other individuals with same point of interest, to gather and share knowledge and first-hand information. Social networking websites act like an online society of users who is familiar with internet.

Social networks are developed with more advance features after year 2003. Since Facebook holds the most number of active users it became referred by the name social network. Figure 1 indicates the leading social networks worldwide as of January 2014, ranked by number of active users.

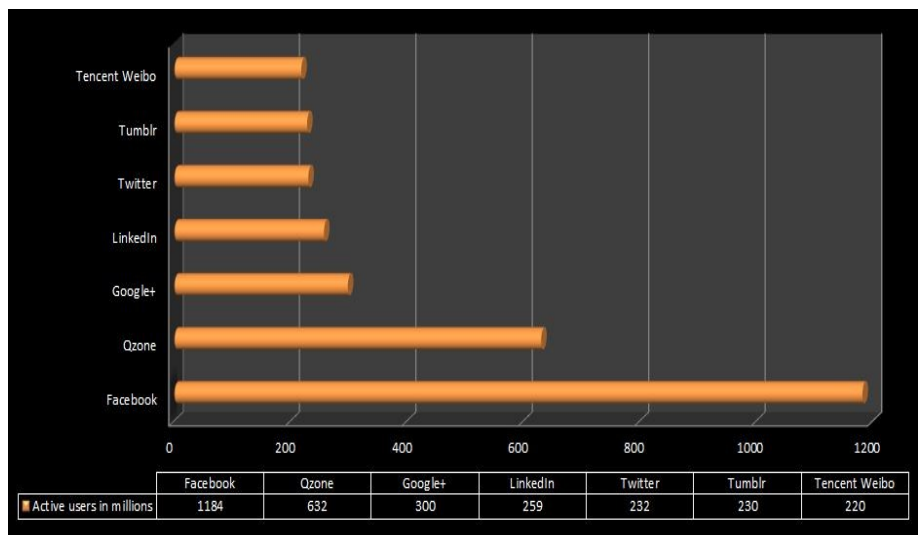

Fig. 1: Active users of top seven social networking sites (source: http://www.statista.com/)

*B. An Overview of Facebook*

The advancement in technology and communication leads to a considerable improvement in social networking such as Facebook and MySpace, used to keep in touch or share information with other individuals. By giving people the control over whole site, we're making the world more transparent (Zuckerberg, 2007). Making the world get connected is the ultimate goal behind Facebook. Facebook is an online directory that connects people through social networks in universities (Zuckerberg, 2005).

Facebook was launched in 2004 by Mark Zuckerberg, Dustin Moskovitz and Chris Hughes to help university students in purpose of identifying students who are residing in other residences. One month later, it was expanded by Mark and friends to any Harvard university students. Later, Facebook extended to all high schools local area networks, and then eventually expanded to internet users all around the world. In 2008 Facebook reached 100 million active users, half of them are spending more than 20 minutes in Facebook site per day (Facebook, 2008). Figure 2 shows the rapid growth of Facebook during the years 2008-2013.

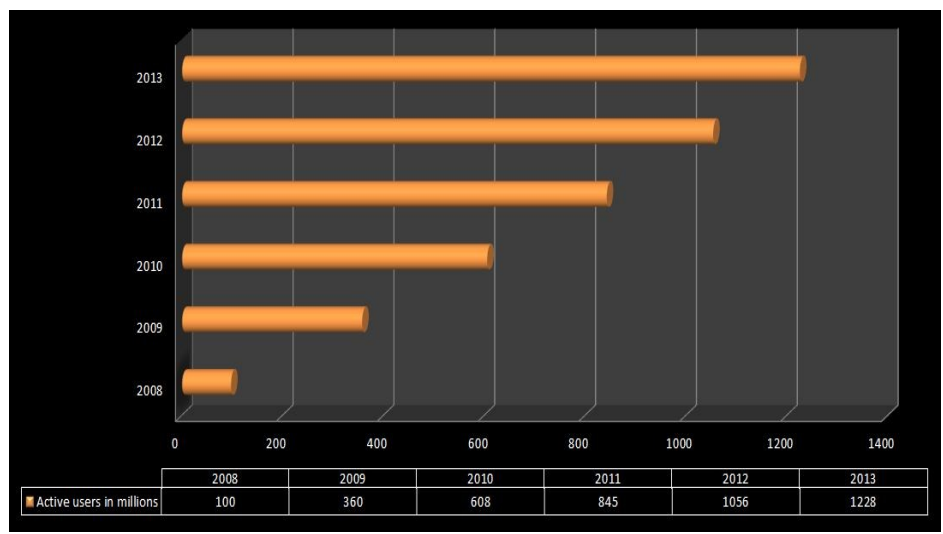

Fig. 2: Active users of Facebook from 2008 to 2013(source: http://www.statista.com/)

As like most of other social networking sites, Facebook is a single-page application(SPA) offers a well-organized web page where users' can store their personal info and make connections with same type of users. The disclosure of friends, not only each user view other's personal profile but also their whole networks. This functionality allows users' to traverse from friends profile to anywhere, so that individual's social network grows rapidly across the world (Walther et al., 2008). This ability or capability is the backbone of Facebook and became the only reason of its rapid growth with comparing other social networking sites. Apart from this the simplicity of Facebook is one significant reason for attracts millions of users around the world.

*C. Sri Lankan University Students Engagement in Facebook*

Facebook is completely a communication tool for users'. Facebook is used by the huge number of undergraduate students and graduate students on a regular basis. There is huge amount of professional and common interest in the effects of social networking on undergraduate student development and achievement (Abramson, 2011). Students use Facebook in various ways to accomplish a wide range of social responsibilities and just for fun too. In university students' perspective the widespread social media website is Facebook, anywhere between 85 and 99% of university students use Facebook (Jones & Fox, 2009).

In 2012 the World Bank reported that the Internet users in Sri Lanka were last reported at 2.5 million in 2010. Among those most of them are teenagers, Figure 3 shows the distribution of Facebook users in different age groups. Sri Lanka has nearly 1.2 million Facebook accounts, 20% of those are fake accounts (Sri Lanka Police, 2014). Even though there is a fair amount of legal accounts in Sri Lanka to consider.

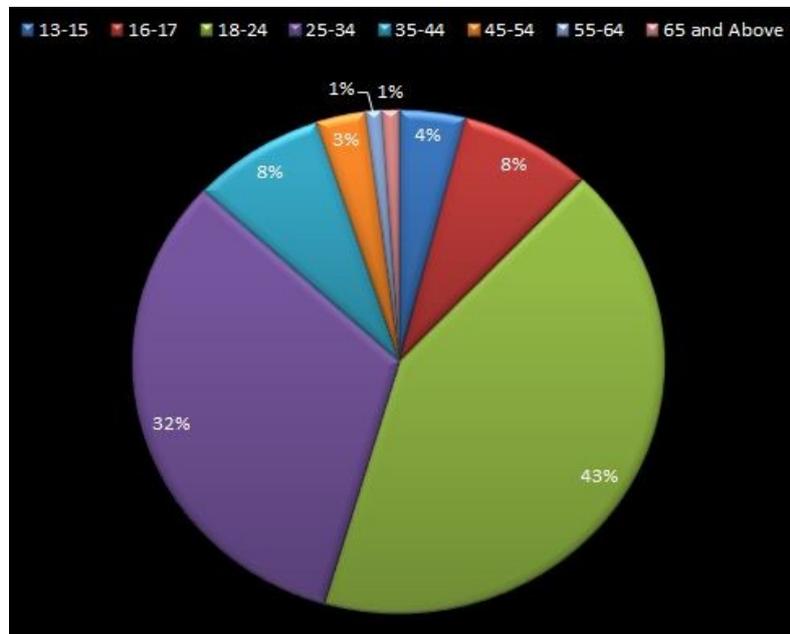

Fig. 3: Active users of Facebook in different age groups, Sri Lanka
(source: http://www.statista.com/)

The above most recent data, collected by the www.statista.com from Sri Lankan Facebook users shows that of the 43% of Facebook users are in between the age 18 and 24. Most users in this group contain university students, Facebook allows them to express themselves, interconnect, and collect profiles that highlight their abilities and capability throughout lifetime.

Social Networking Sites allow students to express themselves, communicate, and collect profiles that highlight their talents and experience. There are numerous reasons demanding young people to use Facebook and even few adults also addicted to it and need to use Facebook. Based on the meeting we had with university students, we found many reasons as stated below the youths want to use Facebook.

- Always make connection with friends
- It is good for time passing
- Helps in studies, a sort of collaborative study
- Makes it very easy to distribute information and good content to batch mates
- Easily publish life events, posts any photos and videos
- Reduce the stress
- Interface of Facebook is uncluttered and clean
- Discovery and explore the interests in both academic and personal interest

D. *Academic Performance of Student*

The students' performance (academic achievement) plays an important role in producing the best quality graduates who will become great leader and manpower for the country thus responsible for the country's economic and social development (Ali et.al, 2009). The undergraduates who obtain high quality and good education can contribute the country hugely. The use of internet and World Wide Web is an important factor imposing the academic performance. Somehow Facebook affects the academic performance of students. Most of the researcher around the word used the GPA to measure the student performance (Galiher, 2006; Darling, 2005; Broh, 2000; Stephen & Schaban, 2002). In Sri Lanka University Grant Commission defined the level of academic performance through the same Grade Point Average point system.

### III. METHODOLOGY

#### A. Information Gathering

Information gathering done through a web based survey among university students. Students who use the Facebook selected for this study. The web questionnaire was designed and distributed to selected students from five Universities across Sri Lanka – Sabaragamuwa University of Sri Lanka, University of Jaffna, University of Peradeniya, Wayamba University of Sri Lanka and University of Kelaniya. To get truthful information the university students were requested to respond the questionnaire anonymously.

Two hundred and eighty seven (287) students participated in this survey. In the sample, only 32.5% students were from non-science streams and others were from science stream. The majority of students are full time course followers (91%).

#### B. Measurement of Variables

The following variables were used to develop and test cases;

i. Time spending on Facebook
   In this study, time spend on Facebook measured as how often he/she visits the Facebook and spends how much time by actively using Facebook. We categorized the frequency and time spending on Facebook in following manner.

| Category | Number of times per day |
|---|---|
| A-1 | Below 2 |
| A-2 | 2-5 |
| A-3 | More than 5 |

Table 1: Categorization of students in terms of frequency of access per day

| Category | Number of hours per day |
|---|---|
| B-1 | Below 2 |
| B-2 | 2-5 |
| B-3 | More than 5 |

Table 2: Categorization of students in terms of number of hours per day

ii. Grade Point Average (GPA)
   It varies from 0.0 to 4.0 depend on the grades of the students in Sri Lankan universities'. Categorization of GPA considered in following way as given in Table 2.

| Category | GPA |
|---|---|
| C-1 | 0-2 |
| **C-2** | 2-3 |
| **C-3** | 3-4 |

Table 3: Categorization of students in terms of GPA

## IV. RESULTS AND DISCUSSIONS

The test cases we defined will inspect the correlation between the frequency or amount of time spent on Facebook and the amount which a student participates in academic activities. Initially we think that there is an inverse relationship between time or frequency with GPA, as the more frequent/time spent on Facebook, the less time a student engage with his/her academic activities. Since majority students falls in to A-2 category, most of the students are moderate users' of Facebook. Table 4 and Table 5 show the distribution of students in each category.

|     | C-1 | C-2 | C-3 |
| --- | --- | --- | --- |
| **A-1** | 17 | 44 | 20 |
| **A-2** | 18 | 101 | 11 |
| **A-3** | 37 | 33 | 6 |

Table 4: GPA distribution of students with frequency of use

|     | C-1 | C-2 | C-3 |
| --- | --- | --- | --- |
| **B-1** | 8 | 90 | 30 |
| **B-2** | 12 | 71 | 5 |
| **B-3** | 52 | 17 | 2 |

Table 5: GPA distribution of students with time spending in Facebook

*A. Test Case 1*

In this case, occasional Facebook users (A-1) analyzed with corresponding GPA categories, prime factor of student academic performance. Category A-1 students are less interested in Facebook; there may be inverse relationship with academic activities.

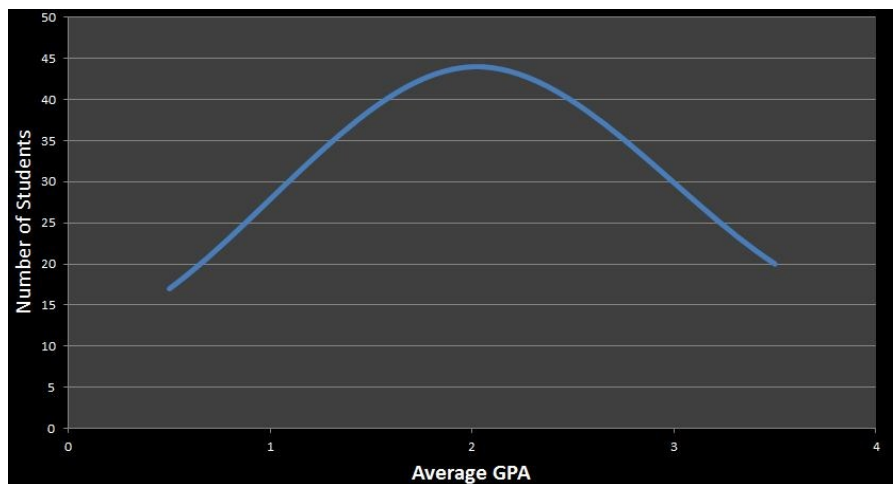

Fig. 4: Distribution of A-1 category users'

Of the occasional Facebook users, most of them received high grades, resides in category C-2 and C3.

*B. Test Case 2*

In this case, medium frequent Facebook users (A-2) analyzed with corresponding GPA categories. Category A-2 students are somehow interested in Facebook but not addicted.

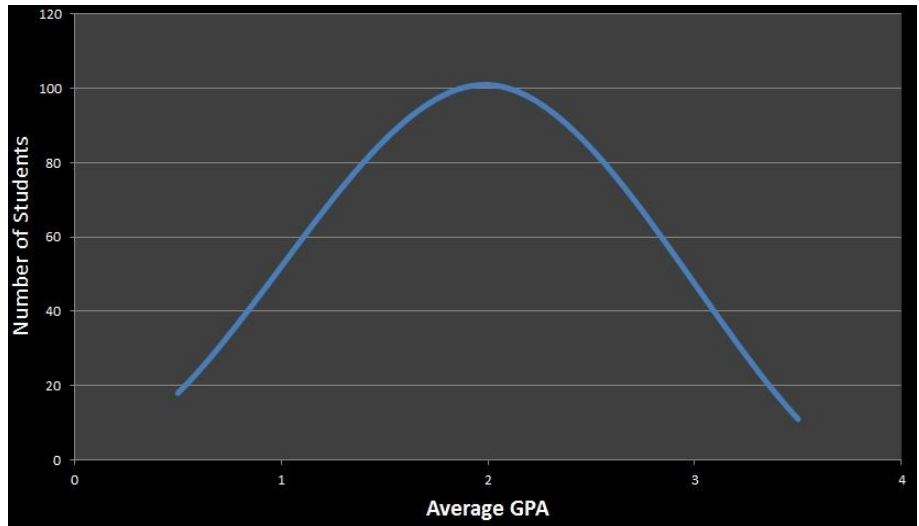
Fig. 5: Distribution of A-2 category users'

Of the medium frequent Facebook users, most of them received medium level grades, resides in category C-2.

*C. Test Case 3*

In this case, frequent Facebook users (A-3) analyzed with corresponding GPA categories. Category A-3 students are who addicted to Facebook.

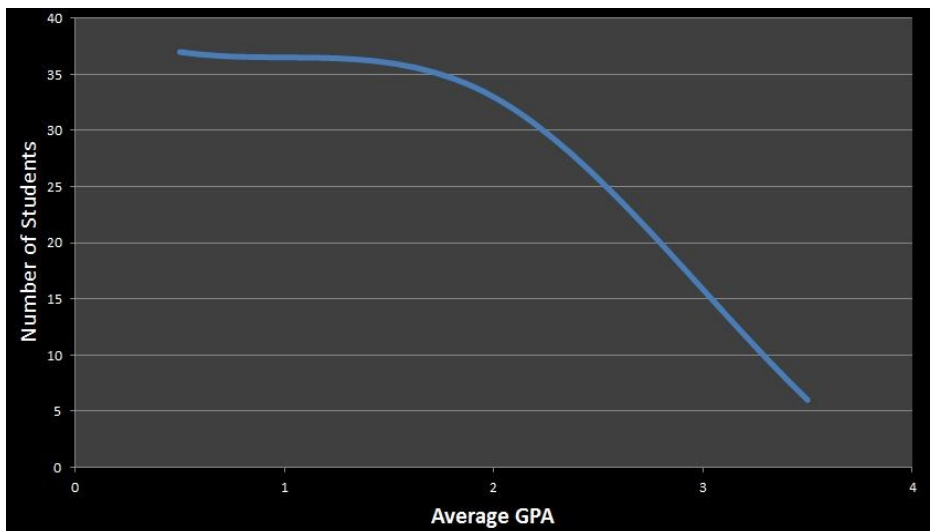
Fig. 6: Distribution of A-3 category users'

Of the frequent Facebook users, most of them received medium grades and low grades, resides in category C-1 and C-2.

*D. Test Case 4*

In this case, light Facebook users (A-4) analyzed with corresponding GPA categories. Category A-4 students are very occasional users of Facebook.

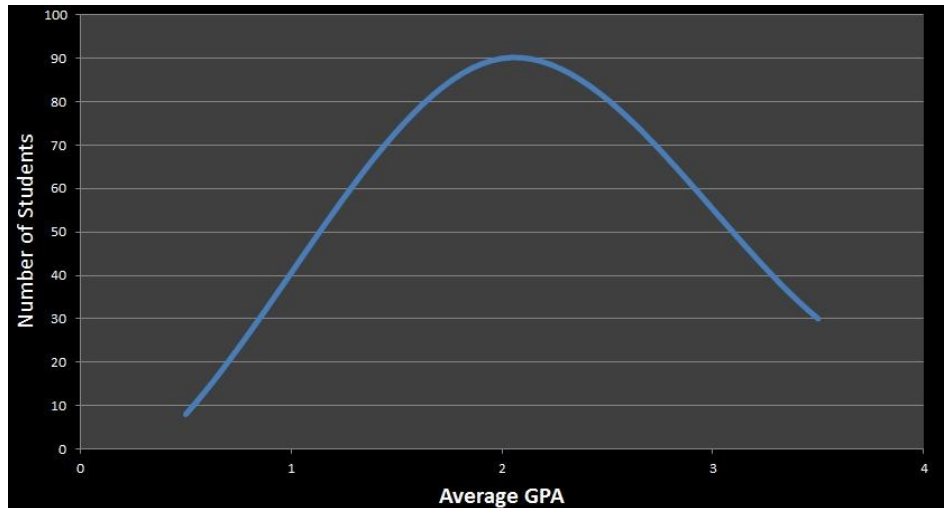
Fig. 7: Distribution of A-4 category users'

Of the light Facebook users, most of them received medium and high grades, resides in category C-2 and C-3.

*E. Test Case 5*

In this case, medium time spending Facebook users (A-5) analyzed with corresponding GPA categories. Category A-5 students are somehow interested to stay in Facebook for some time.

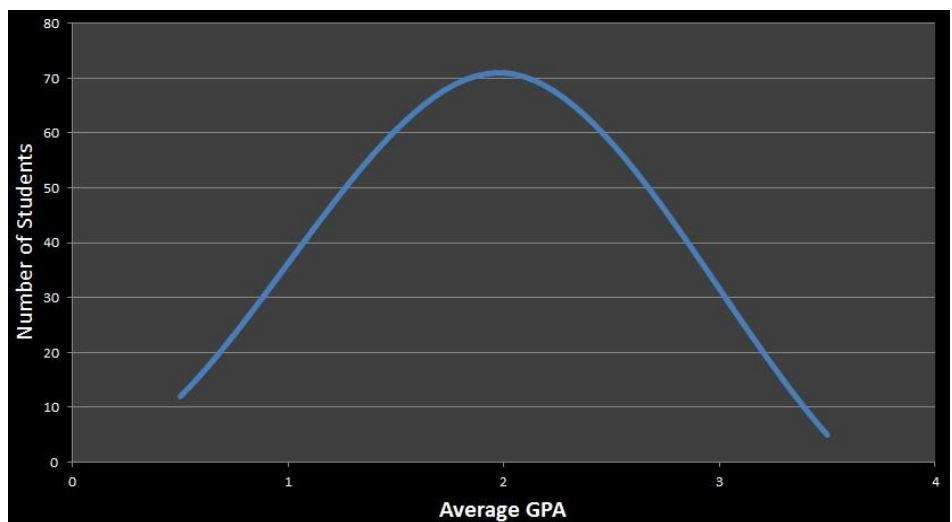
Fig. 8: Distribution of A-5 category users'

Of the medium time spending Facebook users, most of them received medium grades, resides in category C-2.

*F. Test Case 6*

In this case, heavy Facebook users (A-6) analyzed with corresponding GPA categories. Category A-6 students are addicted in Facebook and spend much more time.

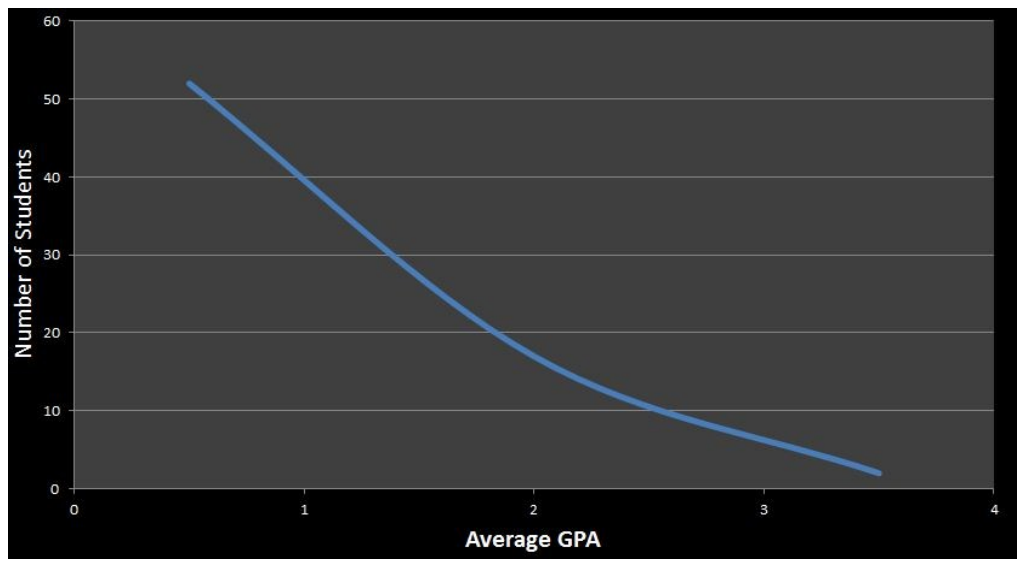

Fig. 9: Distribution of A-6 category users'

Of the heavy Facebook users, most of them received medium and low grades, resides in category C-1.

## V. CONCLUSION

The study found the correlation between social media usage and academic performance. Most of the heavy or frequent users received low grades, compared to light users. We found similar results with lower grades. By considering Test Cases there is a significant difference in Grade Point Average between those considered to be heavy or frequent users of social media and those considered to be light or occasional users. As we employ more time on Facebook, a significant performance decrement should be there. The results of our study indicate that time and the frequency of using Facebook were predictors of academic performance. In addition, it could predict the quality of life as well.

However, an unanticipated finding was that there are numerous positive usages of Facebook still employed. In future we expect to expand the positive usages of Facebook among university students which help them to increase their academic performance.

ACKNOWLEDGMENTS

We would like to thank universities participate in this survey. Extend our sincere thanks to students who spent their precious time and for their assistance and support. Further we would like to acknowledge all friends who assisted us in conducting survey.

Reference: **504374**